\documentstyle[prl,aps,twocolumn,epsf]{revtex}

\newcommand{\beq}{\begin{equation}}
\newcommand{\eeq}{\end{equation}}
\newcommand{\bea}{\begin{eqnarray}}
\newcommand{\eea}{\end{eqnarray}}
\newcommand{\eps}{\epsilon}
\newcommand{\veps}{\varepsilon}

\newcommand{\benn}{\begin{displaymath}}
\newcommand{\eenn}{\end{displaymath}}

\begin{document}

\title{ Casimir Interaction among Objects Immersed 
in a Fermionic Environment }

\author{Aurel Bulgac$^1$ and Andreas Wirzba$^2$}

\address{ $^1$ Department of Physics, University of Washington,
Seattle, WA 98195--1560, USA\\
$^2$ Institut f\"ur Kernphysik (Theorie),
Forschungszentrum J\"ulich, 
D--52425 J\"ulich, GERMANY        }

\date{\today }
\maketitle
\begin{abstract}
Using ensembles of two, three and four spheres immersed in a
fermionic background we evaluate the (integrated) density of states
and the Casimir energy. We thus infer that for sufficiently smooth
objects, whose various geometric characteristic lengths are larger
then the Fermi wave length one can use the simplest semiclassical
approximation (the contribution due shortest periodic orbits only) 
to evaluate the Casimir energy. We also show that the Casimir energy
for several objects can be represented fairly accurately as a sum of
pairwise Casimir interactions between pairs of objects.  
\end{abstract}

\draft




\narrowtext

In 1948 Casimir predicted the existence of a very peculiar effect, the
attraction between two metallic parallel plates in vacuum
\cite{casimir}. The existence of such an attraction has been confirmed
experimentally with high accuracy only recently \cite{lamoreaux}. The
origin of this attractive force can be traced back to the modification
of the spectrum of zero point fluctuations of the electromagnetic
field. Similar phenomena are expected to exists for various other
(typically bosonic) fields \cite{others,kardar} and the corresponding
forces are referred to as Casimir or fluctuating interactions. A
related interaction arises when the space is filled with
(noninteracting) fermions, which is particularly relevant to the
physics of neutron stars \cite{nm,earlier} and quark gluon plasma
\cite{qgp}. Spin--orbit interaction will be neglected as well.
One of the simplest cases corresponds to nuclei embedded
in a neutron gas. These however could be replaced with buckyballs
immersed in an electron gas, in liquid mercury for example.
Particularly attractive candidates for the study of this type of
Casimir effects in essentially perfect degenerate fermi systems are
the dilute atomic Fermi condensates \cite{condensates}.

In the case of two parallel impenetrable planes, dimensional arguments
suggest that the dependence of the Casimir energy for fermions on the
distance between the two planes has the form
\beq
E_C = \mu F(k_Fd),
\eeq
where $\mu = \hbar ^2 k_F^2/2m$ is the chemical potential, $k_F$ is
the Fermi wave vector and $d$ is the distance between the two planes.
For this simple geometry it is straightforward to evaluate the
function $F(k_Fd)$ \cite{nm}. One has to be careful and specify
whether the calculation should be performed at fixed particle number
or fixed chemical potential, as one can easily show that the Casimir
energy has a different behavior in these two limits.

For more complicated geometries the evaluation of the Casimir energy
is generally a rather involved, even though straightforward, numerical
procedure.  Our main goal is to reach a qualitative understanding of
the Casimir energy in the case of complicated geometries.  We shall
consider mainly two obvious limits, when the objects immersed in the
Fermi environment are either much smaller or much larger than the
Fermi wave length. The limit of small scatterers is relatively easy to
treat and is considered mostly for the completeness of the
analysis. We show here that the case of large scatterers can be
treated quite accurately using semiclassical methods at practically
all separations. The most important conclusion we are able to draw
from our study however is that the Casimir interaction energy in the
case of more than two scatterers can be evaluated quite accurately as
a sum of pairwise interactions between these scatterers. This
conclusion comes to some extent as a surprise, since it is known that
Casimir energy is not pairwise additive, in other words, the
interaction among extended objects cannot be evaluated as a sum of
pairwise interactions, see e.g. Ref. \cite{others,kardar}.

Let us consider at first the case of two impenetrable spheres of
radius $a$ at a distance $r\ge 2a$ between their centers. In order to
calculate the Casimir energy we shall represent the sufficiently
smoothed fermion density of states (smoothing is over an energy
interval larger than the level spacing in the 
volume $V$ of the
entire system):
\beq
g(\varepsilon,a,r)=g_0(\varepsilon)+g_W(\varepsilon,a) +
                   g_C(\varepsilon,a,r) ,
\eeq
where $g(\varepsilon,a,r)$ is the total fermion density of states,
$g_0(\varepsilon)$ is the density of states in the absence of
scatterers (ideal Fermi gas), $g_W(\varepsilon ,a)$ is the correction
to the density of states arising from the presence of two spheres
infinitely apart from each other and $g_C(\veps,a,r)$ is the remaining
part, which is of central interest to us here (in the following we
shall not make explicit the $a$-- or $r$--dependence, but show only
the energy  ($\varepsilon$--) dependence).

In the case of $N$ scatterers 
the Krein formula \cite{krein,uhlenbeck}
provides a link between the $N$--body scattering matrix $S_N(\veps )$
and the change in the 
density of states due to the presence of $N$
scatterers, namely
\beq
\delta g(\veps) = g(\veps)-g_0(\veps) 
= \frac{1}{2\pi i} \;\frac{d \ln \det S_N(\veps )}{d \veps }.
\eeq
Following Refs. \cite{andreas}, the determinant of the
$S_N(\veps)$--matrix can be represented as follows
\bea
& & \delta g(\veps) 
= \frac{1}{2\pi i} \;\frac{d \ln \det S_N(\veps )}{d \veps }  
= g_W(\veps ) + g_C(\veps ) \\
& &= \frac{1}{2\pi i}
\frac{d }{d \veps } \ln
\left [ \prod _{j=1}^N \det S_1(j,\veps )\right ] + 
\frac{1}{2\pi i}
\frac{d }{d \veps } \ln
\left [  \frac{\det M^\dagger (\veps ^*) }{\det M(\veps )}\right ]
, \nonumber
\eea
where $ M(\veps )$ is a Koringa--Kohn--Rostoker (KKR) multiple
scattering matrix \cite{kkr}. $ g_W(\veps )$ determines the change in
the density of states due to the presence of isolated scatterers,
which in case of large scatterers is given basically by a Weyl
formula, see Refs. \cite{hilf,brack} for various examples and general
formulas. $g_C(\veps )$, which is determined by the multiple
scattering KKR--matrix $M(\veps )$, vanishes in the limit of
infinitely separated scatterers and is the only part of the density
of states which depends on the relative arrangement of the
scatterers. The Casimir energy at fixed particle number can then be
introduced as
\bea
& & E_C = 
\int _{-\infty} ^\mu \veps g(\veps )d\veps -
\int _{-\infty} ^{\mu_0} \veps [g_0(\veps )+g_W(\veps)]d\veps  
                      \label{eq:cas}\\
& & \approx
 \int _{-\infty} ^{\mu _0} (\varepsilon -\mu _0 ) g_C(\veps ) d\veps   
= -\int  _{-\infty} ^{\mu_0} {\cal{N}}_C(\veps ) d\veps
,\\
& & {\cal{N}}  =  \int _{-\infty} ^\mu  g(\veps )d\veps =
\int _{-\infty} ^{\mu_0}  [g_0(\veps )+g_W(\veps)] d\veps ,\\
& & {\cal{N}}_C(\veps ) =
\int   _{-\infty} ^\veps d\eps g_C(\eps ) ,
\eea
where the omitted terms are  ${\cal{O}}(V^{-1})$.
${\cal{N}}$ is the total number of fermions, $\mu$ and $\mu_0$
are the values for the chemical potential with the scatterers at
finite and infinite separation from each other respectively and
${\cal{N}}_C(\veps )$ is the relevant correction to the integrated
density of states. Strictly speaking, the quantities $g(\veps)$ and
$g_{0}(\veps )$ are infinite, as they are proportional to the volume
$V$ of the entire space. This redundant divergence can be handled
easily by considering first a very big box, the volume of which is
subsequently taken to infinity. One can then show that $E_C$ has a
well defined and finite value in this limit.

\begin{figure*}[h,t,b]

\begin{center}
\epsfxsize=6.5cm
\centerline{\epsffile{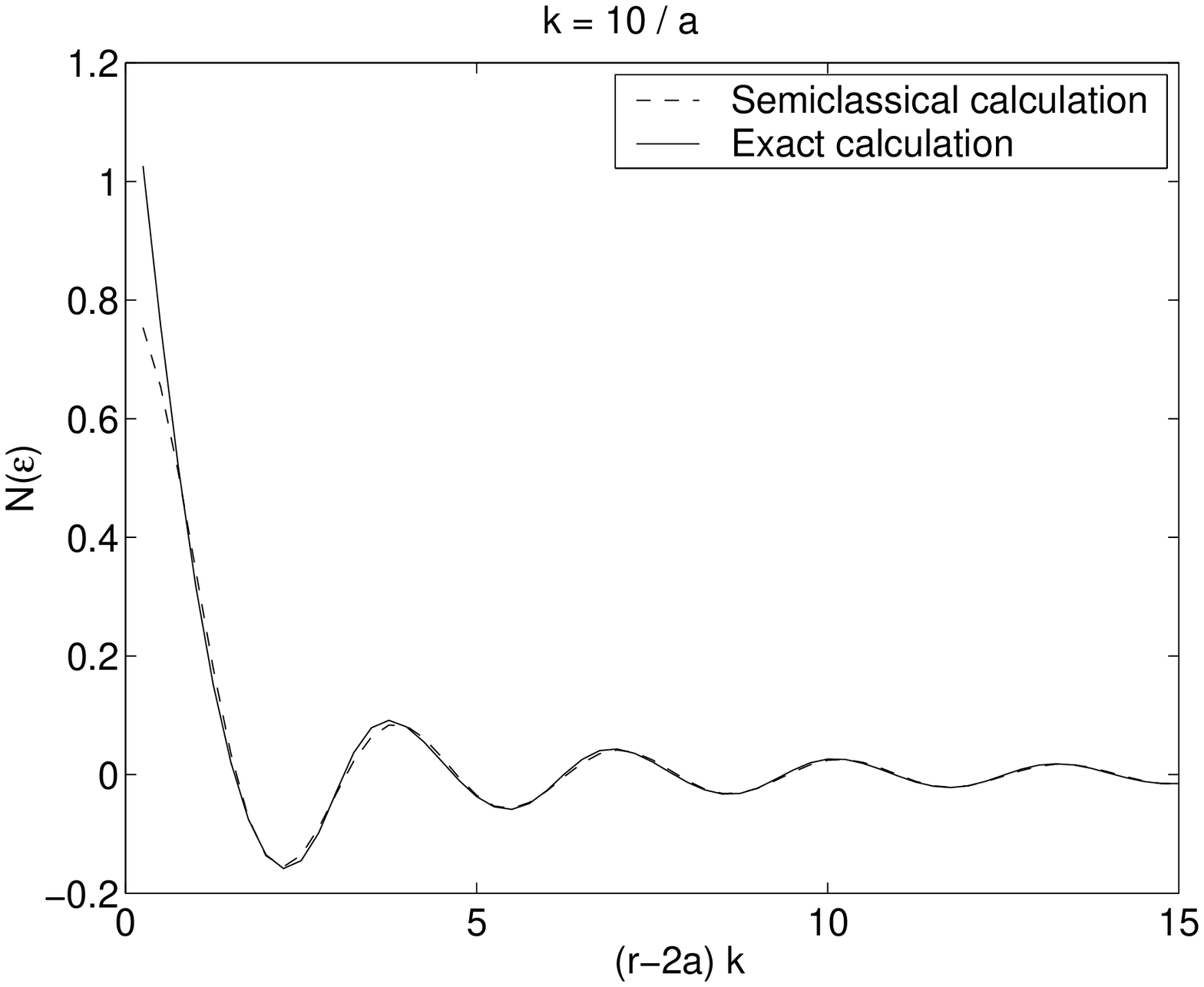}}
\end{center}

\begin{center}
\epsfxsize=6.5cm
\centerline{\epsffile{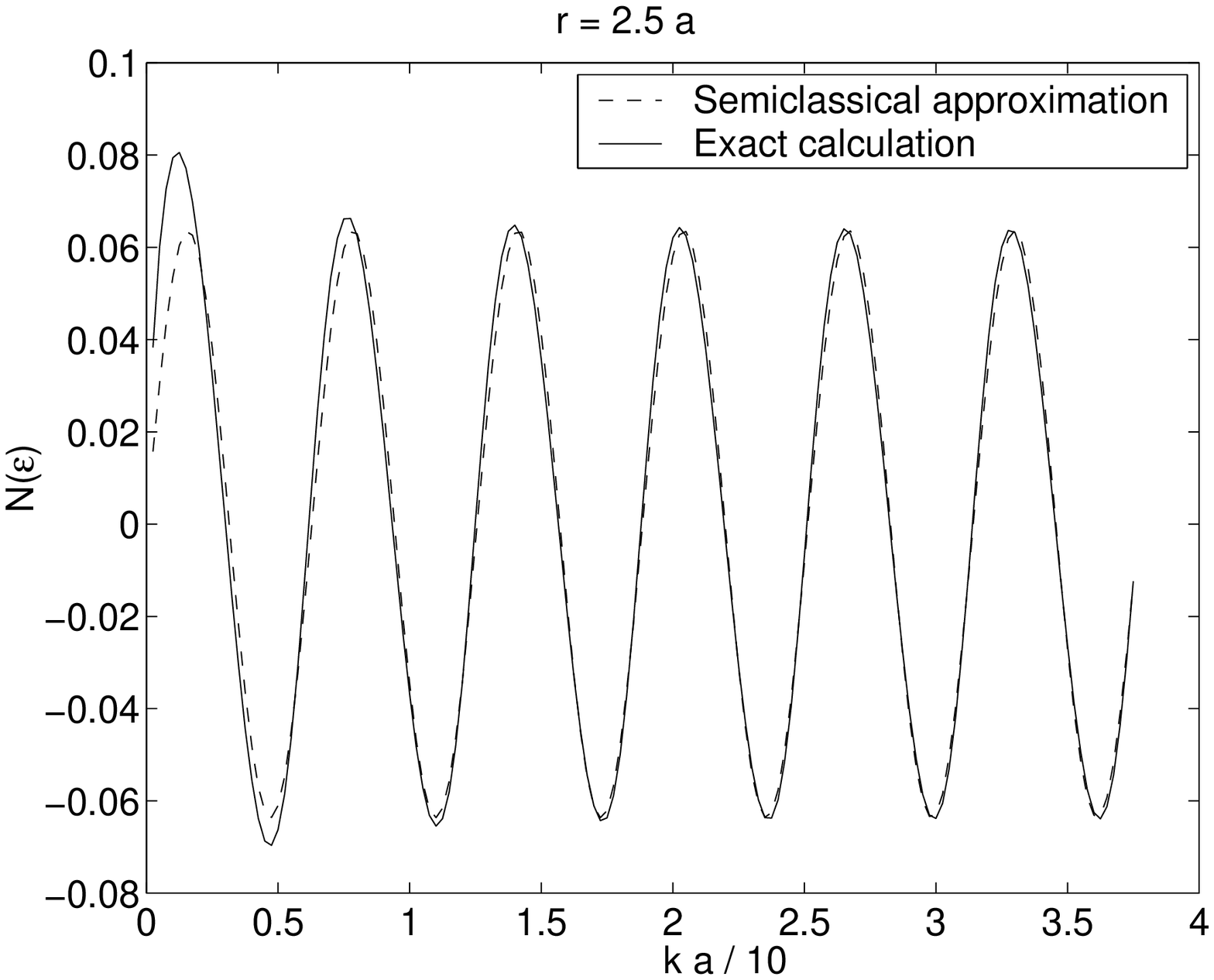}}
\end{center}

\caption{ The correction to the number of states $N_C(\veps )$ as a
function of $r-2a$ in the upper part and as a function of $k$  in the
lower part for $\nu =1$}

\label{fig:fig1}

\end{figure*}

Using the explicit formulas for the KKR--matrix from
Ref. \cite{andreas}, one can compute numerically $g_C(\veps )$ for
various arrangements of hard spherical (or circular in 2D) scatterers.
It is possible to obtain significantly simpler expressions for the
(integrated) density of states in the limit of very small and very
large scatterers. If the wave length $\lambda =2\pi/k$ ($\veps = \hbar
^2 k^2/2m$) is much larger than the 
radii of the scatterers and the scatterers do not overlap 
then one can show that the KKR--matrix $M(\veps)$ is given by 
(see \cite{RWW96} for the analog in the 2D case)
\beq
[ M (\veps)]_{nm}  \approx
\delta _{nm} - (1-\delta _{nm}) f_n(\veps)
 \frac{ \exp (ikr_{nm})}{r_{nm}}  , \label{eq:kkr0}
\eeq
where the indices $n,m=1,\dots,N$ run over the scatterers, $r_{nm}$ is
the distance between the centers of the $n$--th and $m$--th
scatterers, $f_n(\veps)$ is the $s$--wave scattering amplitude on the
$n$--th scatterer. 
In the case of two 
spheres of radius $a$, with their centers $r$ apart ($r\gg a$) one obtains 
\beq
{\cal{N}}_C(\veps )= \nu
\frac{a^2}{\pi r^2} \sin [ 2k(r-a)] + 
{\cal{O}}\left((ka)^3, \frac{a^4}{r^4} \right) 
\label{eq:small} ,
\eeq
where $\nu$ is the spin degeneracy factor.
The next order correction arises from $p$--wave scattering. In the
case of a finite radius $a$, one can use the 
Gutzwiller trace formula
to determine this correction to the (integrated) density of states
semiclassically ($scl$) 
\cite{brack}
\bea
& &\delta g_{scl}(\veps) =\nu 
\sum_{po}\frac{(-1)^{m_{po}} 
T_{ppo} }{\pi\hbar \sqrt{|\det (M_{po}\mbox{$-$}1)| } }
\cos \left (
\frac{S_{po}}{\hbar} \right ) , \\
& &{\cal{N}}_{scl}(\veps) = \nu
\sum_{po}\frac{ (-1)^{m_{po}} }{n\pi\sqrt{|\det (M_{po}\mbox{$-$}1)| } }
\sin \left (
\frac{S_{po}}{\hbar} \right ) ,
\label{eq:gutz}   
\eea
where the summation is over periodic orbits ($po$), $T_{ppo}$ and $n$ are the
period  and number of repetitions 
of the primitive periodic orbit ($ppo$), $M_{po}$, $S_{po}$ and $m_{po}$
are the stability matrix, classical action and Maslov index
(which counts the number of bounces  under Dirichlet 
boundary conditions) of the $po$ \cite{brack}.
Taking into account only the contribution
arising from the single $po$  of length $2(r-2a)$, with no
repetitions, one derives the following result 
\beq
{\cal{N}}_C(\veps) 
\approx \nu \frac{a^2}{4\pi r(r-2a)} \sin [2k( r - 2a )] \label{eq:ppo}.
\eeq
At large distances ($r\gg a$) the leading term in both cases (small
and large scatterers) has the same analytical structure, apart from an
overall numerical factor.  The $kr\gg 1$ limit of Rel.\,(\ref{eq:ppo})
can be reproduced from  the KKR--matrix in the case of large spheres
($ka>1$) as well. One can expect that the semiclassical result
(\ref{eq:ppo}) is reasonably accurate when the reduced action along the
classical $po$ is larger than unity, i.e. when
$S_{po}/\hbar=2k( r - 2a )> 1$, and that this approximation should
fail when the two spheres are very close.  Surprisingly however, at
smallest separations when
$ r-2a\ll a $ the semiclassical estimate 
is only about 30\%
lower than the exact result. For $2k( r - 2a )> 1$ the semiclassical
expression (\ref{eq:ppo}) is very close to the exact numerical result
obtained using the KKR--matrix, if the spheres are large ($ka > 1$),
see Fig.\,\ref{fig:fig1}. When $ka\gg 1$ a large number of partial
waves contribute, which renders thus the semiclassical limit
valid. One can show, using arguments along the lines in
Ref. \cite{vattay}, that the contribution arising from creeping orbits
are exponentially suppressed, which is intuitively expected. The
contributions arising from repetitions of the primitive periodic orbit
are relatively small, because the long orbits and the repetitions of a
primitive orbit are less stable in 3D than in 2D. For this reason the
simplest semiclassical approximation for the density of states, and
consequently for the Casimir energy, are more accurate for spheres
than for cylinders.

Our findings suggest that for more complicated geometries, if the
curvature radii are larger than the wave length, one can safely
evaluate the density of states using the contributions arising from
short primitive periodic orbits only, without any repetitions. If the
curvature radii are smaller than the wave length then the alternative
simple approximation of small scatterers could be used. Using
Eq. (\ref{eq:ppo}) one can now easily derive an approximate, but
rather accurate, expression for the Casimir energy for two large
spheres ($k_Fa>1$)
\beq
E_C \approx -\nu \mu \frac{a^2}{2\pi r(r-2a)} j_1[2k_F(r-2a)],  
\label{eq:ecas}
\eeq
where $j_1(x)$ is the spherical Bessel function.

\noindent
Under the same approximations the Casimir energy of a 
large sphere at a distance $r$ from an infinite plane reads
\beq
E_C \approx -\nu \mu \frac{a}{2\pi (r-a)} j_1[2k_F(r-a)] .
\eeq
One can naturally expect that if a standing wave with the Fermi
momentum could be formed in between the two spheres then the total
energy of the system is at a (local) minimum, which thus explains the
oscillatory character of this interaction.

\begin{figure*}[h,t,b]

\begin{center}
\epsfxsize=6.5cm
\centerline{\epsffile{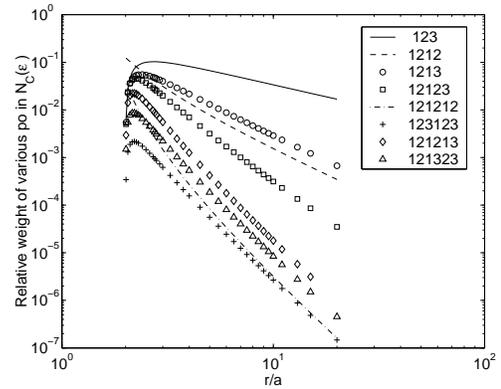}}
\end{center}

\caption{ The relative amplitude of the contribution to the
semiclassical density of states (\ref{eq:gutz}) of $n$--bounce
periodic orbits for the system of three identical spheres, situated at
the vertices of an equilateral triangle, as a function of the distance
between two spheres $r/a (r\ge 2a)$ as compared to the amplitude of the
2--bounce orbit, including the corresponding degeneracies. In the
legend 123 denotes an orbit starting at the sphere 1, followed by a
bounce off the sphere 2, then off the sphere 3 and ending on sphere 1,
and so forth. }

\label{fig:fig2}

\end{figure*}

\begin{figure*}[h,t,b]

\begin{center}
\epsfxsize=6.5cm
\centerline{\epsffile{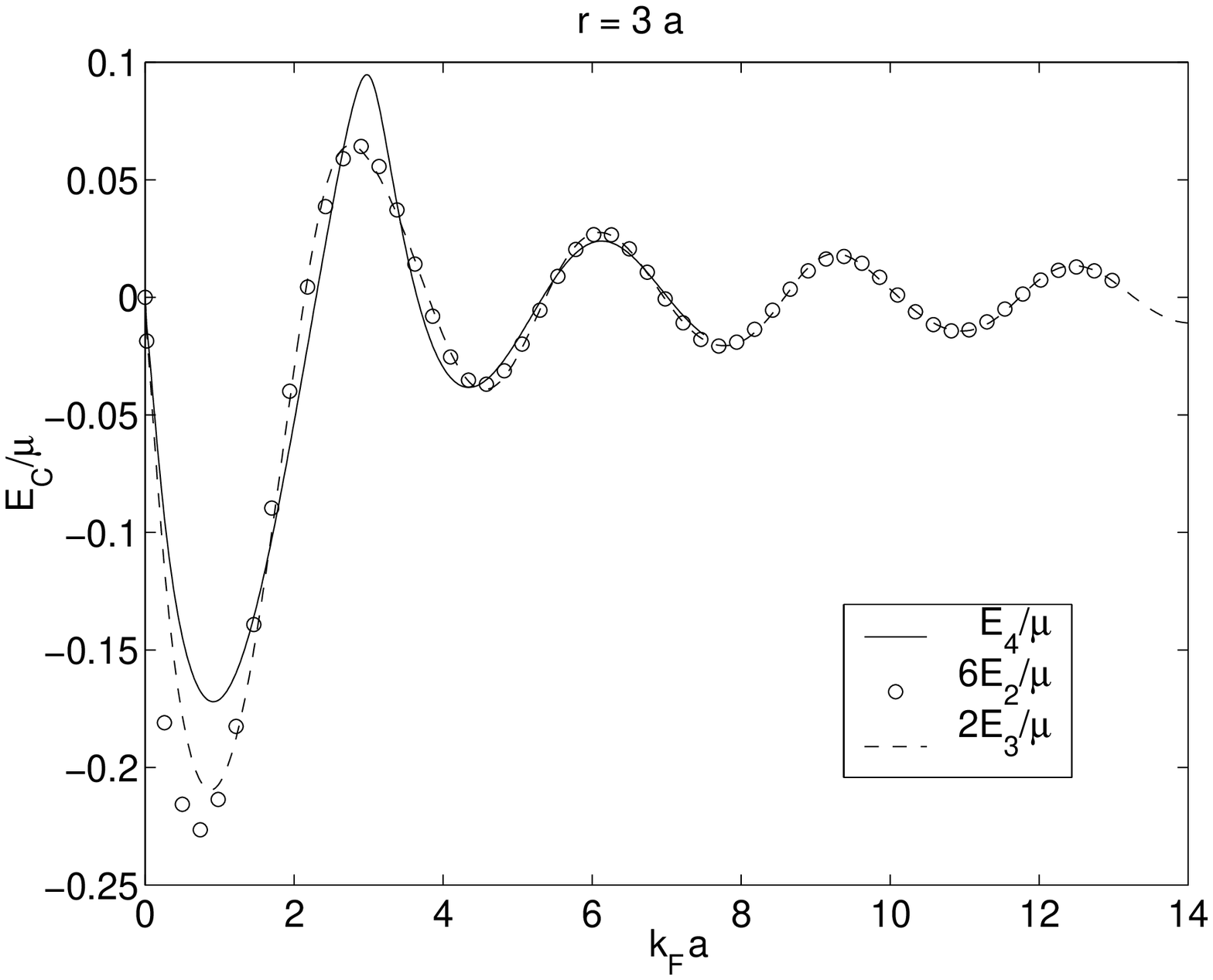}}
\end{center}

\begin{center}
\epsfxsize=6.5cm
\centerline{\epsffile{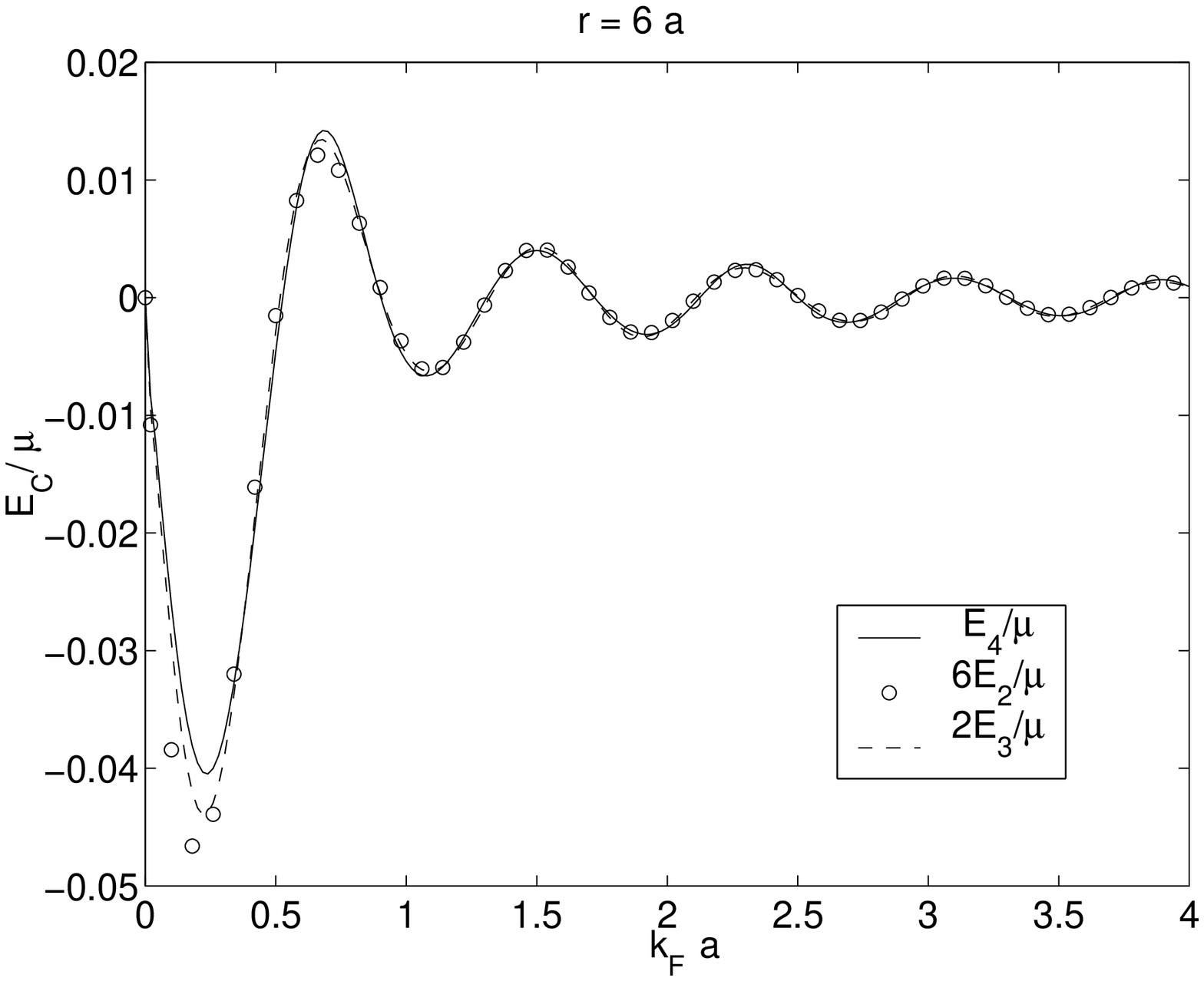}}
\end{center}

\caption{ The ratio of the exact Casimir energy and chemical potential 
$E_C/\mu $ for four spheres  
and computed as a sum of the exact
contributions due to pairs or triplets of spheres for two different
separations. }

\label{fig:fig3}

\end{figure*}

Let us consider now the case of three and four spheres. The semiclassical
picture makes particularly transparent the reason why, strictly speaking, the
Casimir energy cannot be represented as a sum of pairwise interactions. Each
primitive periodic orbit and its repetitions give rise to an additive
contribution to the density of states, see Eq.(\ref{eq:gutz}), and thus to the
Casimir energy (\ref{eq:cas}). For three or more objects there are periodic
trajectories (or standing waves) bouncing off three or more such objects and
thus the contribution to the density of states and to the Casimir energy due
to such orbits depends on the relative arrangement of three or more objects,
thus leading to genuine three and more body interactions.  We determined
however that the contribution of three or more bounce orbits to the density of
states, and thus to the Casimir energy as well, is never dominant.  (NB Our
analysis and conclusions refer to the global domain of the system always and
{\em not} to the fundamental domain or to the one--dimensional representations
of a symmetry reduced problem.) An analysis of the stability matrix of an
$n$--bounce orbit shows that its contribution to the integrated density of
states at large distances is proportional to $1/L^n$, where $L$ is the length
of the orbit, if all the legs of the orbit are comparable in length, see
Fig.~(\ref{fig:fig2}). Any person who ever played pool (thus in 2D) knows
instinctively that long shots are more difficult than short ones and that the
most difficult shots are the many--bounce shots. In 3D and higher dimensions
orbits are typically more unstable than in 2D. An exact evaluation of the
stability matrix for various periodic trajectories shows that even at small
separations the contributions of 2--bounce periodic orbits dominate over those
of three or more bounce periodic orbits. The 3--bounce orbit gives the largest
contribution, an approximately 10 \% corrections, at $r\approx 2.5 a$.  As one
can also judge from Figs. (\ref{fig:fig3}) the role of the orbits bouncing
among three or more objects is never too large. The Casimir energy for three
identical spheres (at the vertices of an equilateral triangle) satisfies the
approximate relation $E_3\approx 3E_2$ and correspondingly in the case of four
identical spheres (at the vertices of a tetrahedron) one has $E_4\approx
6E_2\approx 2E_3$, where $E_N$ stands for the Casimir energy of $N$ spheres,
with high accuracy if $k_Fa\gg 1$. For $k_Fa\le 1$ corrections could reach
10\% for three and 25 \% for four spheres.

We presented here results only for the symmetric arrangement of the
spheres due to the lack of space. Various asymmetrical configurations
of three and four spheres show the same general pattern, namely that
the correction to the integrated density of states
${\cal{N}}_C(\veps)$ can be represented fairly accurately as a sum of
the corresponding corrections for pairs of spheres. Obviously, one can
replace the spheres with other objects, with curvature radii larger
than the Fermi wave length. The pairwise additivity of the Casimir
interaction is reasonably well satisfied as well for the case of point
scatterers, as one can easily check using Eq. (\ref{eq:kkr0}).  We thus
conclude that genuine many--body Casimir interactions are relatively
short ranged and that two--body interactions strongly dominate
-- even for small separations.

We thank Piotr Magierski for discussions. This work has been supported 
financially partially by DoE.

\end{document}